\documentclass{article}
\usepackage{float}
\usepackage{lmodern} 
\usepackage[T1]{fontenc}
\usepackage{authblk}
\usepackage{amsmath}
\usepackage{amsfonts} 
\usepackage{amssymb} 
\usepackage[margin=0.5in]{geometry}
\usepackage[utf8]{inputenc}
\usepackage{graphicx}
\usepackage[backend=biber]{biblatex}

\usepackage{algorithm}
\usepackage{algorithmic}

\bibliography{bibli}

\title{Envelope Estimation using Geometric Properties of a Discrete Real Signal}
\author[1]{Carlos Henrique Tarjano Santos}
\author[2]{Valdecy Pereira}
\affil[1]{(corresponding author, carlostarjano@id.uff.br) Department of Production Engineering, Universidade Federal Fluminense, Rua Passo da Pátria, 156, Campus Praia Vermelha, Bloco D - sala 309, São Domingos, Niterói, RJ, Brasil, CEP: 24.210-240}
\affil[2]{Department of Production Engineering, Universidade Federal Fluminense, Rua Passo da Pátria, 156, Campus Praia Vermelha, Bloco D - sala 309, São Domingos, Niterói, RJ, Brasil, CEP: 24.210-240}

\begin{document}

\maketitle

\begin{abstract}
Despite being an elusive concept, the temporal amplitude envelope of a signal is essential for its complete characterization, being the primary information-carrying medium in spoken voice and telecommunications, for example. Intuitively, the temporal envelope can be understood as a slowly varying function that multiplies the signal, being responsible for its outer shape. 
Envelope detection techniques have applications in areas like medicine, sound classification and synthesis, seismology and speech recognition.
Nevertheless, a general approach to digital envelope detection of signals with rich spectral content doesn't exist, as most methods involve manual intervention, in the form of filter design, smoothing, and other specific design choices, based on prior knowledge of the signals under investigation.
To address this problem, we propose an algorithm that uses intrinsic characteristics of a signal to estimate its envelope, eliminating the necessity of parameter tuning.
The approach here described draws inspiration from geometric concepts to estimate the temporal envelope of an arbitrary signal; specifically, a new measure of discrete curvature is used to obtain the average radius of curvature of a discrete wave, that will serve as a threshold to identify the wave's samples that are part of the envelope. 
The algorithm compares favourably with classic envelope detection techniques based on smoothing, filtering and the Hilbert Transform, besides being physically plausible.
We provide visualizations of the envelope extracted via the algorithm for various real-world signals, with very diverse characteristics, such as voice, spoken and sang, and pitched and non-pitched musical instruments, and discuss some approaches to assess the quality of the obtained envelopes.
A Python module implementing the algorithm was made available via the Python Package Index (PyPI); interactive visualizations of envelopes for a diverse range of digital waves, as well as the source code for the Python implementation, are available online at \textcite{2020TarjanoEnvelope}.
Besides the most direct applications of this work to audio classification and synthesis, we foresee impact in compression techniques and machine learning approaches to audio. We briefly discuss some potential paths in this direction. The discrete curvature definition presented could also be extended to three-dimensional settings, to improve shape detection algorithms based on alpha shapes.
\end{abstract}


\section{Introduction}
Envelope detection, also known as demodulation, is ubiquitous in both analogue and digital signal processing \parencite{2011CaetanoImproved}. Nevertheless, the literature in this area is fragmented \parencite{2017LyonsDigital}. Besides, most envelope detection techniques are designed to account for very specific settings, like pure sinusoids with moderate noise content, a limitation that excludes most physical signals, as is the case of recorded sound, for example; that limitation arises in part due to the lack of a strict mathematical definition of a temporal envelope \parencite{2013Mengempirical}.

In many contexts, however, the temporal amplitude envelope of a signal plays a prominent role: according to \textcite{2017QiRelative}, for example, the envelope is at least as important as the fine structure of a sound wave in the context of the intelligibility of Mandarin tones. This is also the case for the English language \parencite{1995ShannonSpeech}, where even envelopes modulating mostly noise are still capable of conveying meaning. 

The envelope also helps to impart emotion and identity to the human voice \parencite{2018ZhuContributions} and music: Concert halls with envelope preserving characteristics are deemed more pleasant \parencite{2011LokkiEngaging}, for example.

When dealing with broadband signals, approaches tailored to specific applications are prevalent, such as the one presented by \textcite{2014YangFast} for the distributed monitoring of fibre optic or the one formulated by \textcite{2018AssefModeling} in the context of medical ultrasound imaging.

The problem of envelope detection, normally associated with the field of digital signal processing, can be made equivalent to the geometric problem of defining the (outer) shape of a set of points in $ \mathbb{R}^2 $, provided a method for addressing the difference of units in the horizontal, time related, and vertical, amplitude related, axes of a digital signal is available, rendering the original DSP problem readily approachable via geometric techniques.

One direct way of defining the shape of a set of points in $ \mathbb{R}^2 $ is using the concept of a convex hull: as defined in \textcite{2008BergComputational}, the convex hull of a set of points $ S $ in two-dimensional space is the smallest convex subset of the plane that contains all the points in $ S $. Convex hulls are unique, in the sense that a set of points define one and only one convex hull.

Generalizing that definition, \textcite{1983Edelsbrunnershape} introduced the concept of alpha shapes, that can be seen as a concave hull; a mathematically well-defined generalization of the convex hull of a finite set of points, closely related to the Delaunay triangulation and Voronoi diagrams of those points. 

The intuition behind the alpha shapes algorithm is that only points that can be touched by a circle of radius $ \alpha $ coming from an infinite distance towards the set of points of interest are part of the frontier, that is, of the concave hull of this set.

Once a radius $ \alpha $ is defined, alpha shapes conserve the uniqueness property, in fact tending to the convex hull when $ \alpha \to \infty $. Consider, for example, the artificially generated discrete wave in figure \ref{fig:Hulls} below, where the links between consecutive samples were deemphasized to highlight the “set of points” nature of a signal: The convex hull and a concave hull, in the form of an alpha shape with a particular value of $ \alpha $ are illustrated.

\begin{figure}[H]
  \centering
    \includegraphics{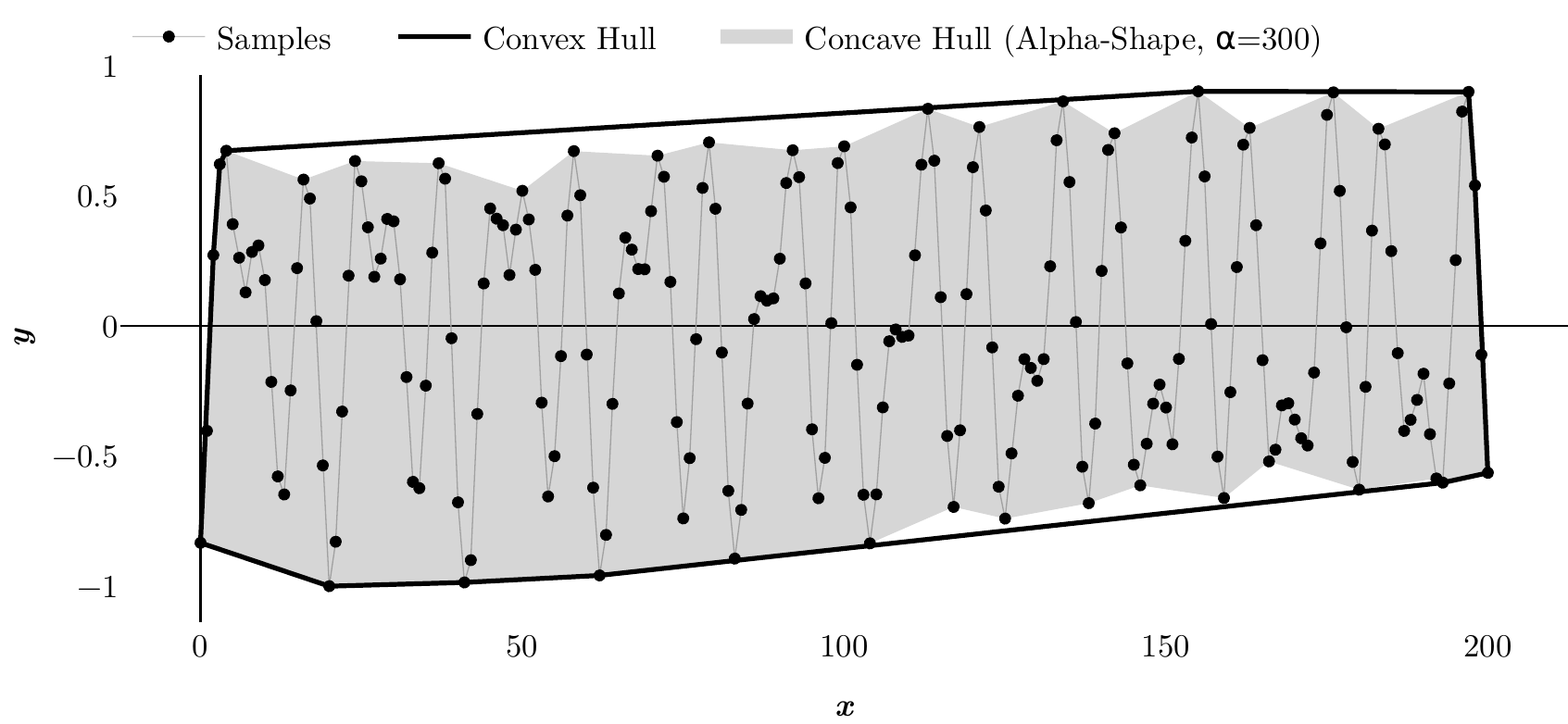}
  \caption{A discrete signal interpreted as a set of points in the Cartesian plane, and the convex and concave hull at $ \alpha=300 $ defined by it.}
  \label{fig:Hulls}
\end{figure}

The alpha shapes approach is used in areas such as detection of features in images \parencite{2016VarytimidisAlpha}, reconstruction of surfaces from a cloud of points \parencite{2015WuAutomated} and spectroscopy \parencite{2019XuModeling}. 

This last work, which involves the estimation and removal of the Blaze function (a kind of envelope) of an echelle spectrograph, being particularly illustrative of the potential synergy between geometric and DSP approaches.

Other steps in the direction of translating geometric algorithms to the context of envelope detection were made by \textcite{2015YangSkeleton} via an algorithm based on the construction of a skeleton underlying the digital wave of interest, and also via the direct translation of computer vision methods to the task of envelope detection \parencite{2015YangRepresenting}.

Following this path, we present a general approach to envelope detection, exploiting the intrinsic characteristics of a generic, spectrally complex wave, in order to completely avoid the need for manual intervention or parameter tuning.

The rest of this paper is divided as follows: We start by defining as specifically as possible what a temporal envelope is, especially in the context of this work and providing a mathematical definition, in terms of the Hilbert transform, for the case of a narrowband signal.

We proceed to illustrate the representation of a discrete wave used in this work, largely based on the concept of dividing the wave into its constituent pulses, and how it can simplify the algorithm, reducing the high dimensionality generally present in digital representations of waves.

Next, we review methods for the estimation of the curvature of a discrete wave, after which we present our own; this method will then be used to estimate the equivalent circle, that will be used to define the envelope of the wave, via the procedure described in sequence.

In the results section, we present some illustrations of the envelopes obtained with our method, comparing both accuracy and running time with the most common algorithms encountered in the literature. We also suggest an alternative use of the algorithm to identify the negative and positive frontiers of a wave, that can be seen as partial envelopes considering only the positive and negative pulses of a wave, respectively.

We follow with suggestions about how the approach here presented can be useful beyond envelope detection, exemplifying with some applications, such as the extractions of the average waveform of a digital wave, and concluding with commentary on potential future developments of the method.

\section{Methods}
After defining what will be considered an envelope in the context of this work, in this section we explore some simplifications that can be applied to a discrete wave, defining the representation that will be used in the subsequent part of the work. 

We next explore the concept of discrete curvature, putting forth our own method, tailored to the definition of the equivalent circle of a wave; we end this section by showing how this circle can be used to identify the envelope of a discrete wave, in a similar approach to that of the alpha shapes theory, but taking advantage of some unique structure present in discrete waves.

\subsection{Characterization of a temporal envelope}

The interest in the theory of envelope detection arose in the analog domain, with the widespread adoption of radio communications, where the problem is typically restricted to simple sinusoids, often with well-defined frequencies \parencite{2011TurnerDemodulation}.

In this context, the most mathematically sound definition of an envelope involves the representation of its underlying wave as an analytic signal, as introduced by Gabor in 1946 \parencite{2007HahnHistory}. In his work, \textcite{1946GaborTheory} applies the then relatively new mathematical machinery of the quantum mechanics to unify the time and frequency-domain representations of a wave, showing how the Hilbert transform could be applied to a real signal in order to obtain an equivalent complex signal, later known as the analytic signal.
This analytic signal has the form $ A(t) = S(t) + \mathbb{H}(S(t)) \ \text{i} $ \parencite{2016HePraat} where $ S(t) $ is the original real signal. $ \mathbb{H}(S(t)) $, the Hilbert transform of the original signal, becomes the imaginary part of the analytic signal; the envelope of a signal thus represented can be straightforwardly obtained by the computation of the complex modulus of $ \mathbb{H}(S(t)) $.

Despite its widespread use, and effectiveness in the context of narrowband signals, envelope detection techniques based on the Hilbert transform don't behave well for broadband signals \parencite{2012DauSpeech}, being even physically paradoxical in some cases \parencite{1996Loughlinamplitude}.

\begin{figure}[H]
  \centering
    \includegraphics{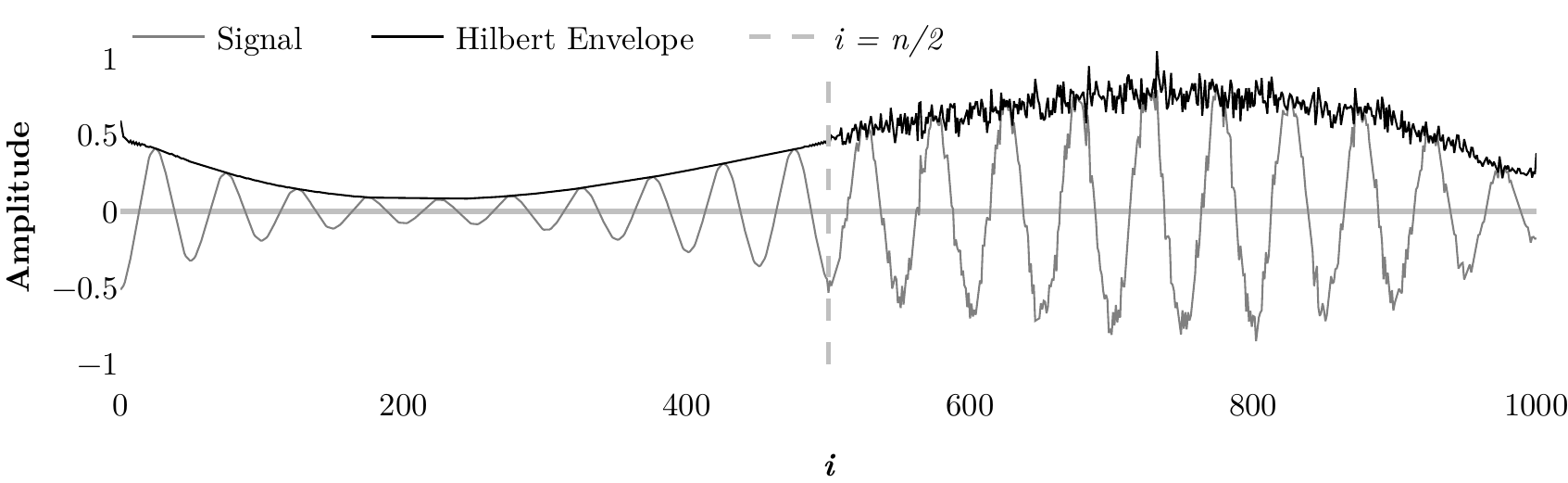}
  \caption{Envelope of a pure sinusoid with a local frequency of 20 cycles modulated by a polynomial of degree 3, as obtained by the Hilbert Transform approach. In the first half, the sinusoid is free of noise, while in the second half Gaussian noise with a standard deviation of $ 1/10 $ of the maximum amplitude of the wave was added to the base signal.}
  \label{fig:AnalyticSignal}
\end{figure}

Figure \ref{fig:AnalyticSignal} exemplifies the Hilbert envelope, for a pure sinusoid modulated by a polynomial, illustrating the difference in the shape of the envelope in the presence and absence of Gaussian white noise. It is readily noticeable from the figure that, as soon as noise is introduced in the original signal, it is reflected in the envelope.

In general, the problem of envelope detection, or demodulation, can be interpreted as the task of, given a continuous wave $ W(t) $, decomposing it in two components $ E(t) $ and $ C(t) $, such that $ W(t) = E(t) C(t) $ \parencite{2011TurnerDemodulation}. 

$ E(t) $ represents the slow varying part of the wave, also known as the (temporal) envelope, modulator component, or amplitude modulation (AM) of $ W $, while $ C(t) $ models its fast varying part, called throughout the literature as its (temporal) fine structure, carrier component, or frequency modulation (FM).

In the case of broadband signals, the problem of envelope detection of an arbitrary wave is ill-posed, in the sense that an infinite number of pairs of $ E(t) $ and $ C(t) $ can result in the same specific $ W(t) $ \parencite{2011TurnerDemodulation, 2013Mengempirical, 1996Loughlinamplitude}.

We address this problem by assuming $ C(t) $ normalized between $ \{-1, 1\}$. This assumption doesn't cause any loss of generality since, given an arbitrary $ C(t) $, we can obtain its normalization by dividing the function by its absolute global maximum, that is, $ \hat{C}(t) = C(t) / \max(|C(t)|) $, provided that $ \max(|C(t)|) \ne 0 $.

In this work we are concerned with the discrete version of this problem: given a finite digital wave represented by the vector $ \textbf{w} \in \mathbb{R}^n $, instead of the continuous function $ W(t) $, obtaining the temporal envelope of $ \textbf{w} $. 

The preceding definitions can be translated to this discrete scenario assuming that the discrete quantities arise from observing the continuous ones at regular time intervals. In that case, the equality $ i = t \ \text{fps} $ can be used to link both settings, where $ i $ is the index of each observation, $ t $ stands for the time in seconds, and fps is the frame rate, or the number of observations made in the period of a second. We can thus define:

\begin{subequations}\label{eq:Envelope}
\begin{align}
  & \mathbf{w}, \mathbf{e}, \mathbf{c} \in \mathbb{R}^n \\
  & \mathbf{w} = (w_0, w_1, \cdots, w_{n-1}) \\
  & \mathbf{e} = (e_0, e_1, \cdots, e_{n-1}) \\  
  & \mathbf{c} = (c_0, c_1, \cdots, c_{n-1}) \\
  & \mathbf{w} = \mathbf{e} \odot \mathbf{c} \ \therefore \ w_i = e_i c_i \ \forall \ i, \ 0 \le i \le n-1
\end{align}
\end{subequations}

\begin{figure}[H]
  \centering
    \includegraphics{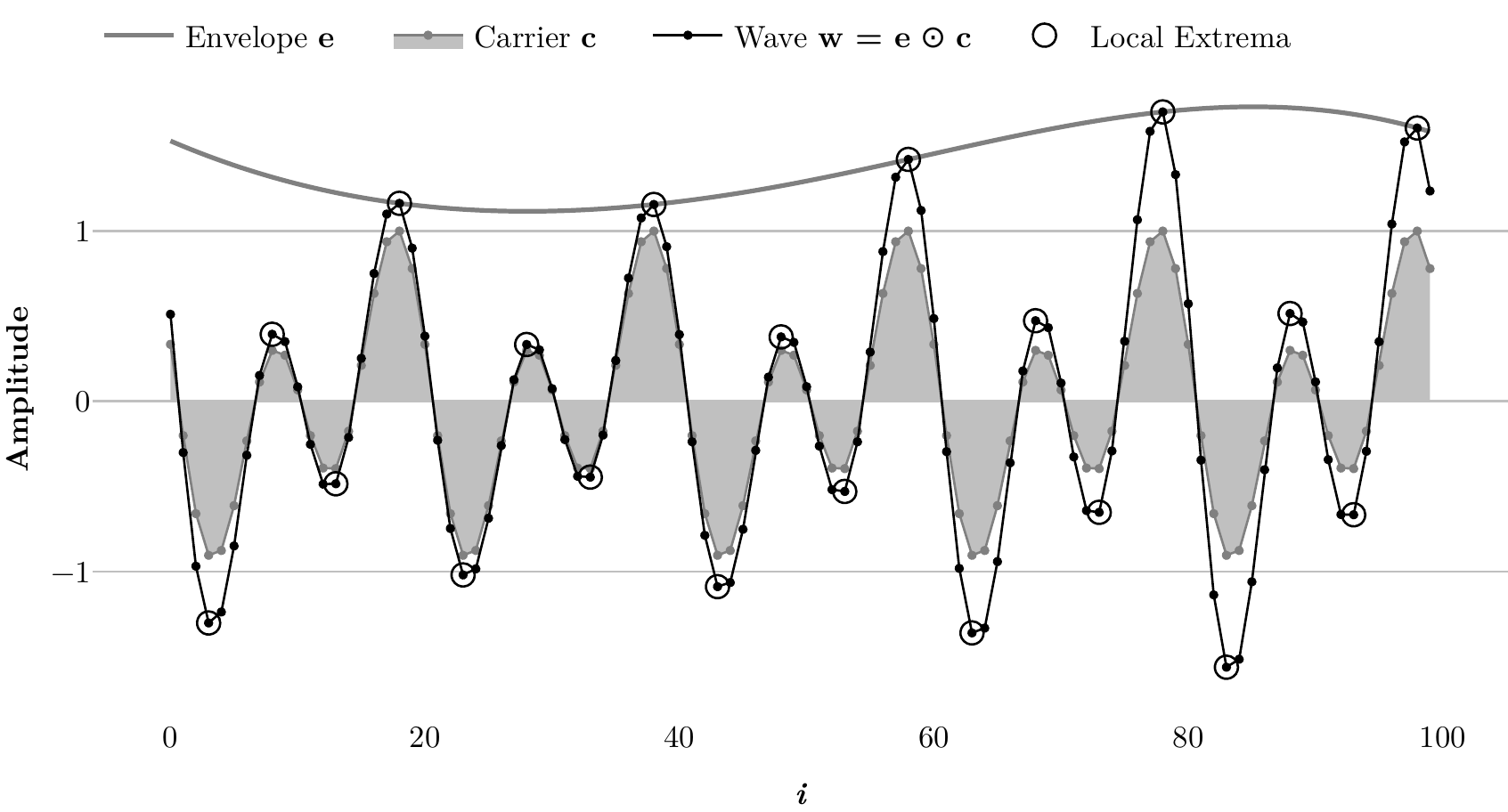}
  \caption{Example of a discrete wave \textbf{w} composed by the element-wise multiplication of an envelope \textbf{e} and a carrier \textbf{c}. The local extrema of \textbf{w} are highlighted with a circle.}
  \label{fig:Envelope}
\end{figure}

The $ \odot $ operator in \ref{eq:Envelope} stands for the Hadamard product, denoting element-wise multiplication of two vectors. Figure \ref{fig:Envelope} provides an example of the vectors just defined: \textbf{w} is obtained by modulating the normalized carrier wave \textbf{c} using the envelope \textbf{e}.

Note from Figure \ref{fig:Envelope} that, at some local maxima, the envelope is well-defined, being in fact equal to the value of the wave $ w_i $, since $ c_i = 1 $ at those points, rendering $ e_i = 1 w_i $. 

The carrier wave \textbf{c} in the figure is periodic, and normalized between $ \{-1, 1\}$. Each cycle of the wave, then, reaches the value 1 only at its maximum making $ w_i $ equal to the value of the envelope at one point in each cycle.

Image \ref{fig:Envelope}, along with the concave hull illustrated in figure \ref{fig:Hulls}, helps forming the intuition behind the method developed in this work: the idea is to identify the local extrema that touch the envelope, those being also the extrema of each (pseudo-)cycle.

This is accomplished with the use of a circle with a defined radius, large enough to prevent it from reaching the smallest local extrema when "rolled" around the (geometric) representation of the wave. Only the extrema touched by the circle will be marked as part of the envelope.

In the next section, a compatible geometric representation will be formulated, along with the necessary tools to estimate the average curvature of the wave and, consequently, the appropriate radius of the circle.

\subsection{Simplified Representation}

Instrumental to the method here presented is the definition of a pulse: each time the value of \textbf{w} changes sign, that is, every time the discrete wave \textbf{w} crosses the horizontal axis, the beginning of a pulse is defined, with the next crossing defining its end. This definition is in line with the one presented in \textcite{national1996telecommunications}, where a pulse is defined as a rapid change in the amplitude of a signal, followed by a fast return to the baseline value; zero, in our case.

From \ref{fig:Envelope} and the discussion in the preceding chapter, we can see that only the maximum point of a positive pulse or the minimum point of a negative one can potentially be equal to the envelope and are, thus, our only points of interest. We can then proceed, for the rest of the method, considering only those points, to great computational economy.

For this reason we define P as the set of the points $ \{P_0, P_1, \cdots, P_{m-1}\} $ where $ P_j = (i_j, \lvert w_j \lvert) $, that is, $ \lvert w_j \lvert $, the absolute value of the extrema of each pulse, becomes the ordinate of each point and $ i_j $, its original index, becomes the abscissa.

\begin{figure}[H]
  \centering
    \includegraphics{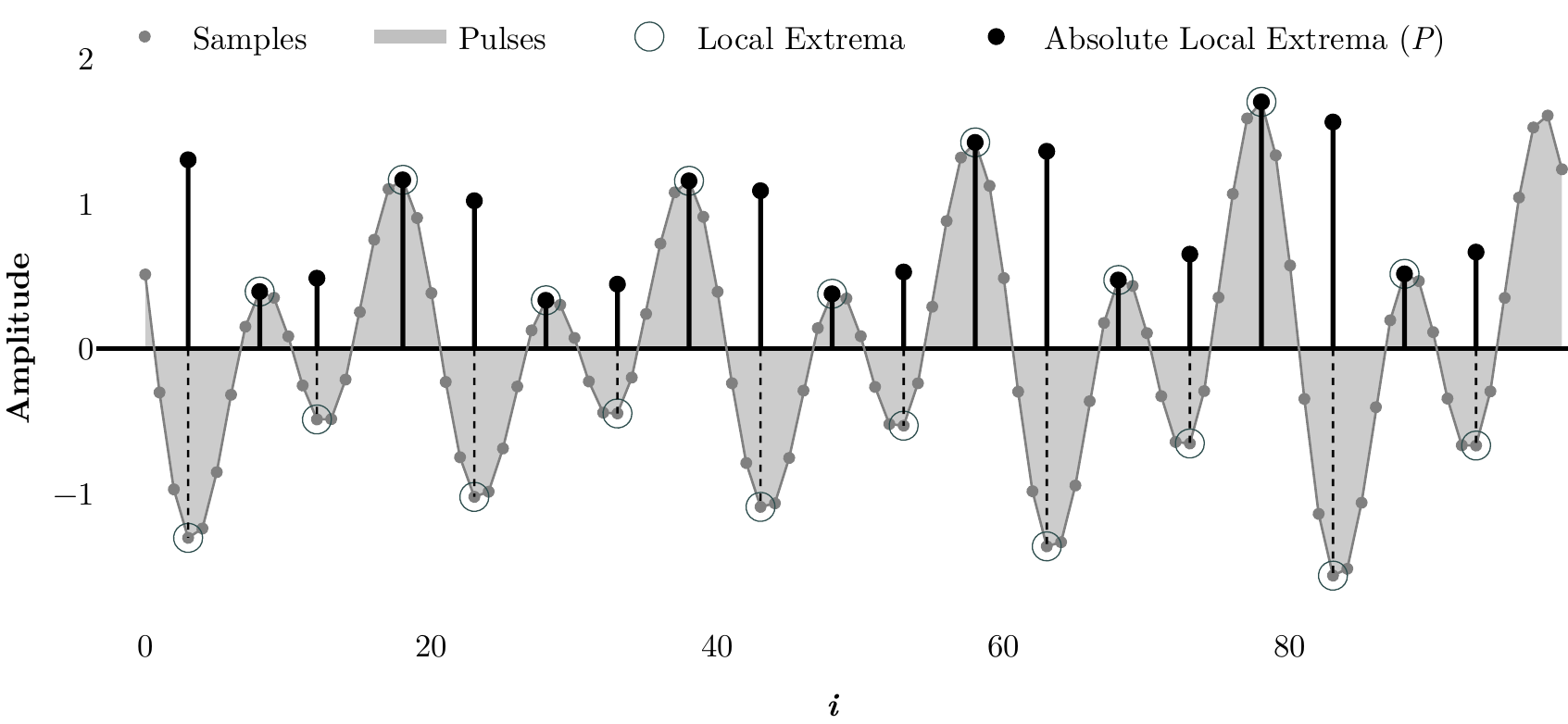}
  \caption{Example of a discrete wave \textbf{w} divided into pulses, of which the extrema are highlighted with a circle. P is the set of the points $ P_j = (i_j, \lvert w_j \lvert) $ representing the absolute value of those extrema.}
  \label{fig:Pulses}
\end{figure}

This relation is illustrated in figure \ref{fig:Pulses}. We call it P to emphasize the connection with the pulses of \textbf{w}.

The points in P are not fit for geometric interpretation yet, since the abscissa and ordinate of their orthogonal coordinate system have different units: in the vertical axis, we have a unit related to the instantaneous intensity of the wave, while in the horizontal axis we have the index at which each extremum occurs, ultimately a time-related unit. 

Specifically, this coordinate system has a basis vector $ \mathbf{b} = \{ \mathbf{b}_x, \mathbf{b}_y \}$ with $ \mathbf{b}_x = (a, 0), \mathbf{b}_y = (0, i) $ where $ a $ is an amplitude unit such as decibel, a measure of the instantaneous sound pressure, volt or even a fraction of a fixed maximum value. $ i $ represents the index of the sample, and is linked to time by the aforementioned relation $ i = t \ \text{fps} $.

We are interested in finding a coordinate system where a pulse's amplitude $ \lvert w_j \lvert $ and length would define, on average, a square. 

To achieve that, we divide the basis vector of the original vertical axis by the average of all ordinates, effectively cancelling the unit of the vertical components of the points. We could divide the basis vector of the horizontal axis by the average of the difference between the abscissa of a point and the abscissa of the immediately posterior, likewise eliminating the axis' unit. 

The drawback is that we would lose the direct link between P and \textbf{w} that arises from the fact that the abscissas of the points in P are indices $ i $ of \textbf{w}: we choose instead to leave the abscissa intact and multiply the now unitless vertical basis axis by the average of the difference between the abscissa of a point and the abscissa of the immediately posterior. In this way, both axes will have unit $ i $, and the relationship is preserved, making it easier to recover the envelope points later.

In this new coordinate system we have a basis vector $ \mathbf{b}^\prime = \{ \mathbf{b}^\prime_x, \mathbf{b}^\prime_y \}$ where $ \mathbf{b}^\prime_x = (i, 0) $ and $\mathbf{b}^\prime_y = (0, i) $, related to the original basis vector by equation \ref{eq:Basis}:

\begin{subequations}  \label{eq:Basis}
\begin{align}
\mathbf{b}^\prime_x &= \mathbf{b}_x  \\
\mathbf{b}^\prime_y &= \frac{\left( \frac{i_{m-1} - i_0}{m-1} \right)}{\left( \frac{\sum_{j=0}^{m-1} \lvert w_j \lvert}{m} \right)} \mathbf{b}_y 
\end{align}
\end{subequations}

\begin{figure}[H]
  \centering
    \includegraphics{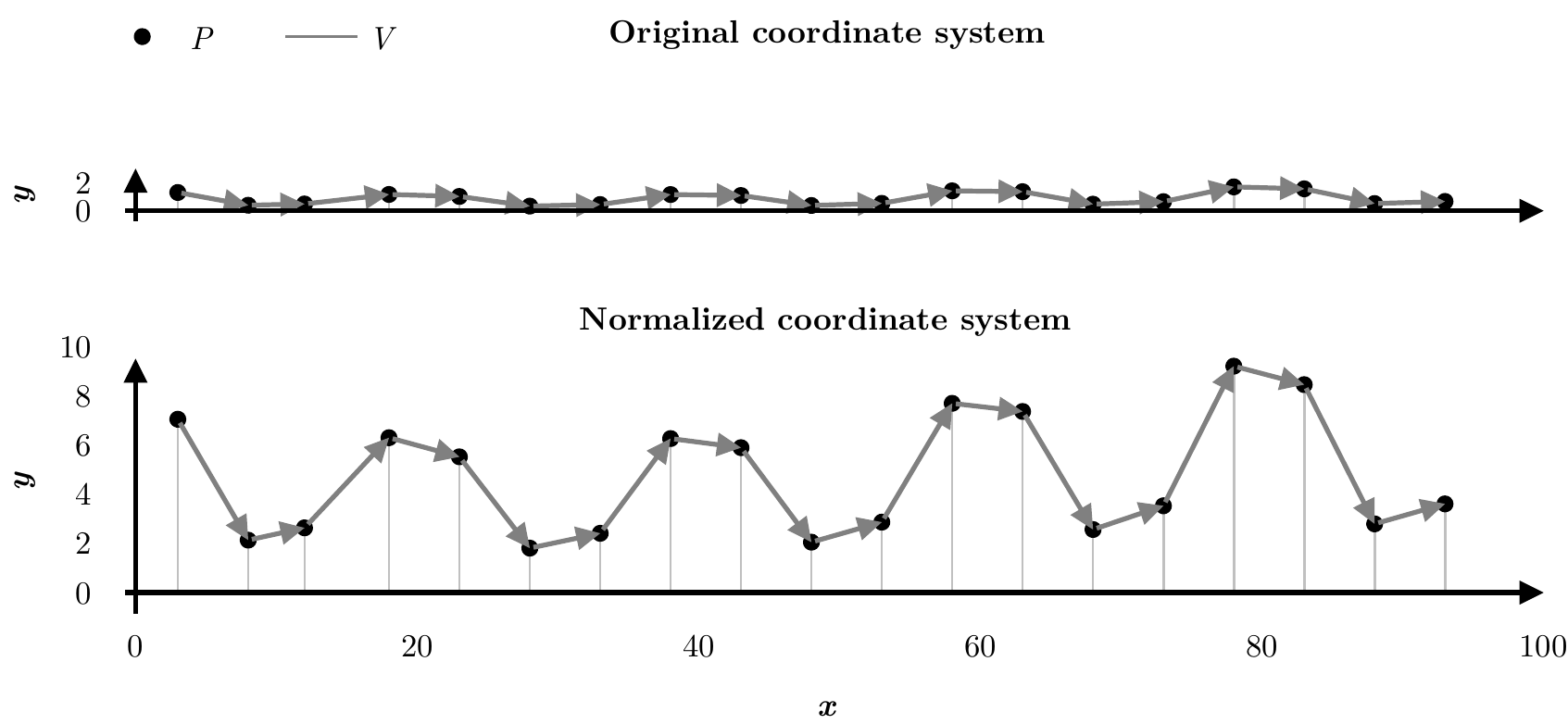}
  \caption{The set of points P and the set V of vectors between two adjacent points of P, in both the original and normalized coordinate systems, shown in scale.}
  \label{fig:Vectors}
\end{figure}

The effect of this normalization can be seen in the points shown in figure \ref{fig:Vectors}, shown in scale in both coordinate systems. The set V of the vectors from each point of P to the next is also shown in the picture, and it will be used to estimate the average curvature of P. We define this set in equation \ref{eq:Vectors} below.

\begin{equation} \label{eq:Vectors}
\text{V} = \{ v_0, v_1, v_{m-2} \} \quad \text{where} \quad v_k = P_{j+1} - P_j \quad \forall \quad 0 \le k \le m-2
\end{equation}

\subsection{Discrete Curvature Estimation}

Translating the explanation of alpha shapes in \textcite{1994EdelsbrunnerThree} to our context, we can say that the points that are part of the envelope are those touched by a circle of a given radius $ r $, coming from $ y \to \infty $ towards the signal, that is not allowed to contain any point of the signal; Intuitively, one can picture a circle being rolled above the signal, and marking the points it touches as envelope points.

To infer the appropriate radius $ r $ of such circle a measure of discrete curvature is needed. Discrete curvature estimation is an important task in image processing \parencite{2010FleischmannNovel} for which no default definition exists. 

The two possible general approaches are the derivation of direct methods, that use characteristics of the discrete wave to calculate the curvature, or the use of the curvature of a smooth, continuous curve fitted to the discrete wave \parencite{2001CoeurjollyDiscrete}.

Concerning direct methods \textcite{2014CarrollSurvey}, for example, derive three such definitions based on the approximation of a circle by an inscribed, centered and circumscribed polygon. In the context of three-dimensional meshes, \textcite{2016VasaMultivariate} evaluate a range of existing estimators from a multivariate point of view.

Those approaches define the discrete curvature in relation to the vertices of a wave (or mesh, in the three-dimensional case). In our particular case, it wouldn't make sense to talk about the curvature of a single pulse, since it is what our points ultimately represent. Instead, we are interested in the change of direction, what the term curvature ultimately means, between two adjacent pulses, as the vectors in the set V suggests.

We thus proceed to define a discrete curvature measure over the edges of a discrete wave. To that end we are going to apply the definition of smooth curvature as the rate of change of the unit tangent to a curve, noting that this is equivalent to that of the osculating circle \parencite{2016VasaMesh}.

\subsubsection{The Equivalent Circle Approach}

\begin{figure}[H]
  \centering
    \includegraphics{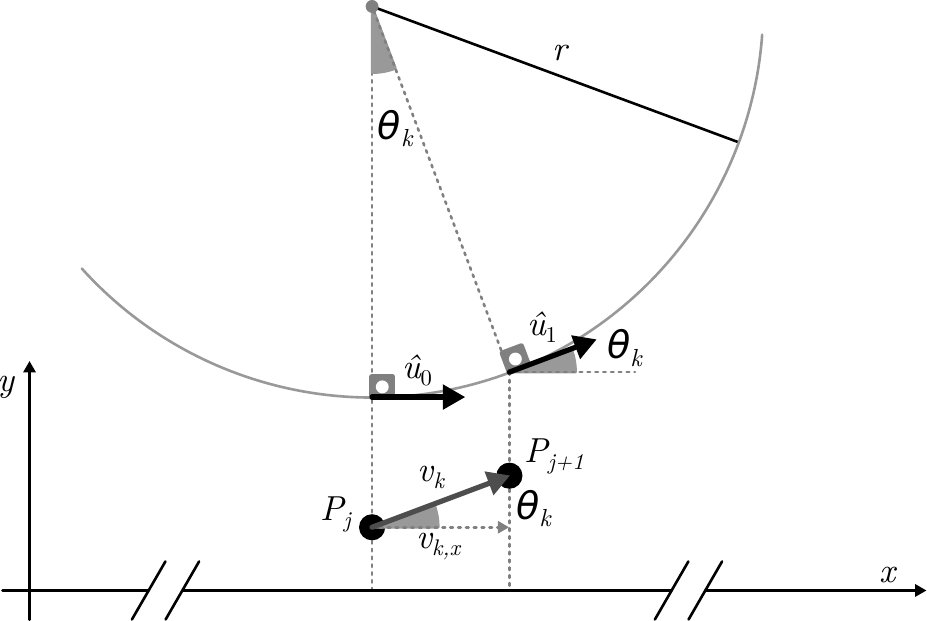}
  \caption{The tangent unit vector of the circle changes from the horizontal direction in $ \hat{u}_0 $ to an inclination of $ \theta_k $ in $ \hat{u}_1 $, $ \theta_k $ being the angle that the vector $ v_k $ makes with the horizontal direction.}
  \label{fig:DiscreteCurvature}
\end{figure}

The rationale is to find, for each vector $ v_k \in \text{V} $, the radius $ r_k $ of the equivalent circle whose tangent has the same change in direction, in the same horizontal distance, as the vector of interest, as shown in figure \ref{fig:DiscreteCurvature}. The average curvature of P will then be obtained by the average of all radii $ r_k $.

From figure \ref{fig:DiscreteCurvature} is easy to see that $ r_k = v_{k,x} \sin(\theta_k)$ where $ v_{k,x} $ is the component of $ v_k $ in the horizontal direction; $ \theta_k $ itself can be obtained using the slope $ m_0 = 0 $ of a line in the horizontal direction and the slope $ m_k = v_{k,y} / v_{k,x} $ defined by $ v_k $ via the equality $ \theta_k = \arctan \big( (m_k - m_0) / (1 + m_k m_0) \big) $.

The radius of the circle that represents the average curvature of V (and P), and will be used to obtain the envelope of the discrete wave \textbf{w} can be obtained via equation \ref{eq:Radius}:

\begin{equation} \label{eq:Radius}
r = \frac{\sum\limits_{k=0}^{m-2} \left( \frac{v_{k,x} \sqrt{v_{k,x}^2 + v_{k,y}^2}}{ v_{k,y}} \right)}{m-2}
\end{equation}

We need now to construct an algorithm, using the radius just obtained, to identify the points that belong to the envelope. Generally, the algorithm for the alpha shapes approach resorts to the construction of the Delaunay triangulation of the set of points, that is latter filtered to contain only the outer edges. We can adopt a more straightforward method, as our points have a defined structure:

Let $ x_o, y_o \in \mathbb{R}, \quad 0 \le y_o < +\infty $ be the coordinates of the center of a circle with radius $ r $, that can be placed anywhere in the upper plane of the coordinate system;

Let $ P_a, P_b, P_c \in \text{P}, \quad P_a \ne P_b \ne P_c $ be three different points of the set P;

For all circles of center $ x_o, y_o $ and radius $ r $ that can be constructed to pass through any two different points of P, that is, $ \forall \quad P_b, P_c $ such that $ (P_{b,x} - x_o)^2 + (P_{b,y} - y_o)^2 = (P_{c,x} - x_o)^2 + (P_{c,y} - y_o)^2 = r^2 $, $ P_a $ will be part of the set $ E $ of the points belonging to the envelope if and only if none of those circles contain $ P_a $, or $ P_a \in \text{E} \leftrightarrow (P_{b,x} - x_o)^2 + (P_{b,y} - y_o)^2 > r^2 $. Summarizing this, we have equation \ref{eq:InEnvelope}:

\begin{subequations} \label{eq:InEnvelope}
\begin{align}
& P_a \in \text{E} \leftrightarrow (P_{b,x} - x_o)^2 + (P_{b,y} - y_o)^2 > r^2, \\
& \forall \ P_b, P_c \mid (P_{b,x} - x_o)^2 + (P_{b,y} - y_o)^2 = (P_{c,x} - x_o)^2 + (P_{c,y} - y_o)^2 = r^2, \\
& P_a, P_b, P_c \in \text{P}, \quad P_a \ne P_b \ne P_c \quad \text{and} \quad x_o, y_o \in \mathbb{R}, \quad 0 \le y_o < +\infty
\end{align}
\end{subequations}

The algorithm follows directly from the definition in \ref{eq:InEnvelope} after noting that the first and last members of P will always be part of the envelope, since they are reachable by a circle of any radius $ r $. From $ P_0 $, our first pivot point, we construct circles passing through $ P_1, P_2, \cdots, P_d $ until a circle that doesn't contain any point in P is found; $ P_d $ then becomes the new pivot point and is included in E, and this procedure continues until $ P_{m-1} $ is reached.

The procedure is formalized in the algorithm \ref{FindEnvelope}. We don't provide similar algorithmic descriptions of the preceding steps since the Python implementation can easily serve as pseudocode.

\begin{algorithm}[H]
\caption{Retrieve Envelope} \label{FindEnvelope}
\begin{algorithmic}
\STATE \textbf{Given} $ \text{P} = \{P_0, P_1, \cdots, P_{m-1}\}, r $, \textbf{circle}, a function that returns the center of a circle of a given radius passing through two points and \textbf{distance}, a function that returns the Euclidean distance between two points,
\STATE \textbf{Let} $ \text{E} = \{ P_0 \}, \text{id1} = 0, \text{id2} = 1 $
\STATE \textbf{Define} $ P_c = \text{point}, \text{empty} = \text{boolean} $
\WHILE {$ \text{id2} < m $}
  \STATE $ P_c \leftarrow \textbf{circle}(r, P_\text{id1}, P_\text{id2}) $
  \STATE empty $ \leftarrow $ \TRUE 
  \FOR{$ i = \text{id2} + 1 $ \TO $ m $}
    \IF {$ \textbf{distance}(P_c, P_i) < r $}
      \STATE empty $ \leftarrow $ \FALSE
      \STATE id2 $ \leftarrow \text{id2} + 1 $
      \STATE \textbf{break}
    \ENDIF
  \ENDFOR
  \IF {empty}
    \STATE E $ \leftarrow $ E $ \cup \ \{ P_\text{id2} \} $
    \STATE id1 $ \leftarrow $ id2
    \STATE id2 $ \leftarrow \text{id2} + 1 $
  \ENDIF
\ENDWHILE
\end{algorithmic}
\end{algorithm}

The set E obtained via algorithm \ref{FindEnvelope} is a set of points. The most straightforward way to transform those points into the vector $ \textbf{e} \in \mathbb{R}^n $ is via linear interpolation, the approach used in this work. 
It is worth noting, however, that many other procedures are available, both for the interpolation and the smooth approximation of points.

\section{Results and Discussion}
Given the lack of a precise definition of what an envelope should be, we begin this section suggesting an empirical metric, based on the behaviour of the carrier wave \textbf{c}. We also comment on some metrics presented in the literature, and compare our algorithm with traditional demodulation approaches, both in relation to execution time and accuracy; to this last end, a numerical indicator is also introduced.

We also illustrate how the method here presented can be used to isolate not only a single envelope of the wave, but the superior and inferior envelopes as well, that we call frontiers, ending the section with a suggestion on how the proposed algorithm can be useful beyond envelope detection, by identifying the approximate locations of the pseudo-cycles in a quasi-periodic wave.

Due to an assortment of factors, such as the unconventional approach presented in this work and the intrinsic difficulty in visualizing discrete waves given their high dimensionality, the authors deemed helpful to provide a companion website with interactive visualizations \parencite{2020TarjanoEnvelope}.

An implementation of the method in pure Python, fully functional, can also be found at the repository dedicated to this work as well as a Python module available at the Python Package Index (PyPI) and conveniently installable via the command `pip install signal-envelope`. Besides, those working under Windows 64-bit can benefit from a specialized dynamic library coded in C++ that is automatically used within this module.

\subsection{Quality of an envelope}

There are no agreed-upon objective metrics to assess the quality of an envelope. Recalling figure \ref{fig:Envelope}, however, and the accompanying discussion, an intuitive way is to consider the carrier wave \textbf{c}, and how well it seems to be bounded by the interval $ \{-1, 1\} $. 

To illustrate how the algorithm performs on such metric, we provide the following figures of original discrete waves \textbf{w}, their extracted envelope \textbf{e}, and the inferred carrier, obtained after dividing, element-wise, the original wave by its envelope.

\begin{figure}[H]
  \centering
    \includegraphics{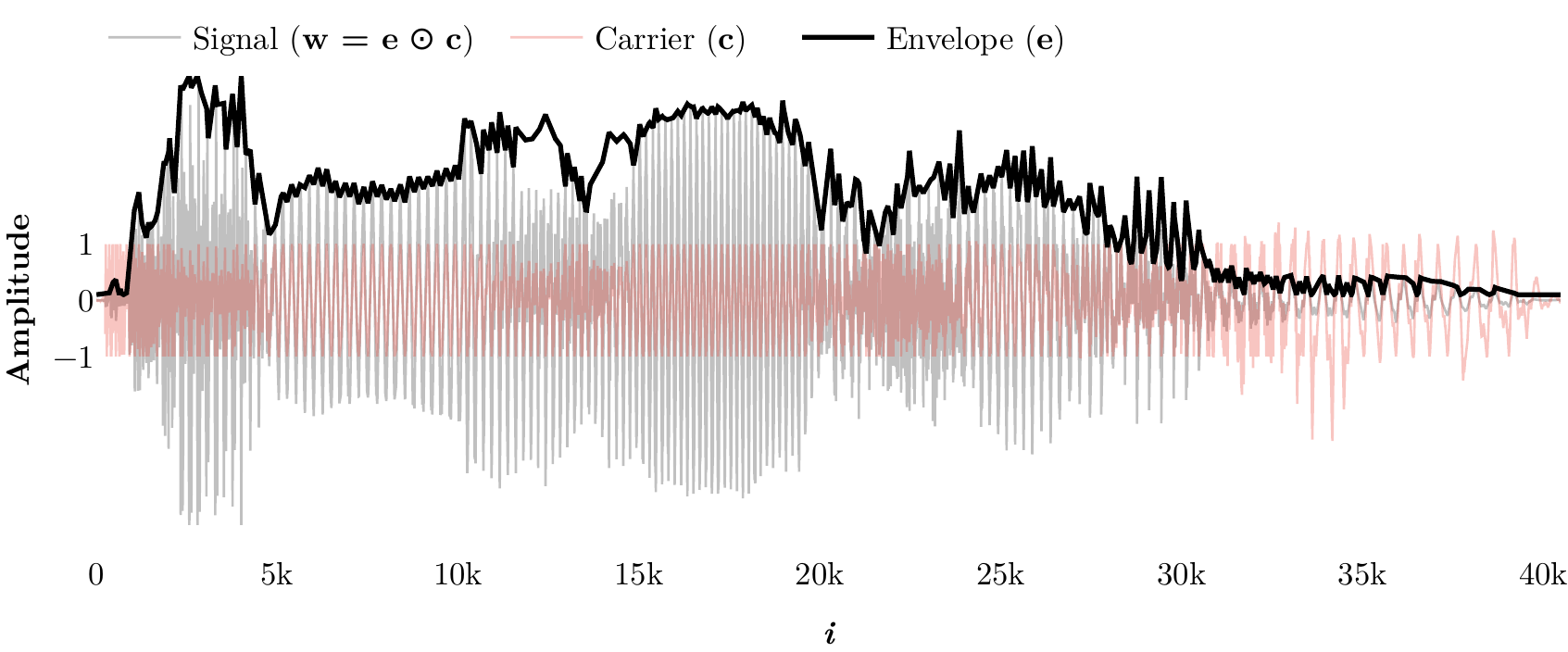}
  \caption{Signal, envelope and carrier for a record of male voice uttering the word “amazing”.}
  \label{fig:spoken_voice}
\end{figure}

From figure \ref{fig:spoken_voice} it can be seen that the algorithm performs well under the carrier metric. This is especially remarkable for the case of spoken voice, a notorious complex wave with fast changes in pitch and high noise content.

\begin{figure}[H]
  \centering
    \includegraphics{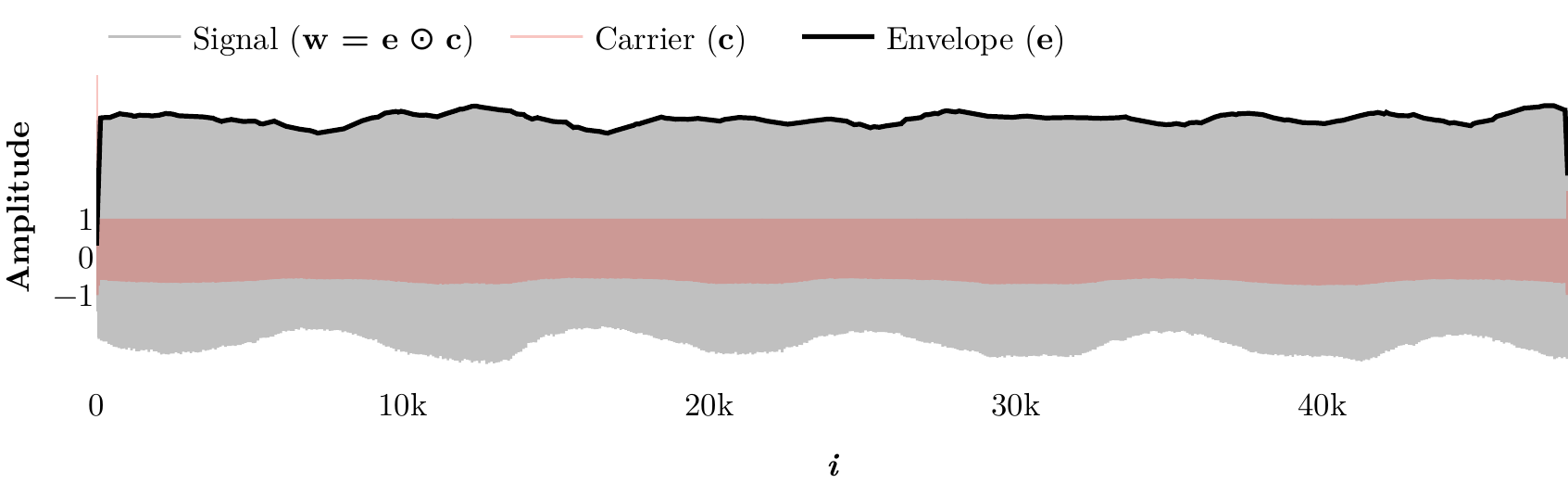}
  \caption{Signal, envelope and carrier for a record of an alto singer.}
  \label{fig:singer_voice}
\end{figure}

These characteristics are in stark contrast with those of the sang voice in figure \ref{fig:singer_voice}, where a recording of an alto singer is shown; there, the wave is almost periodic, with a stable waveform.

\begin{figure}[H]
  \centering
    \includegraphics{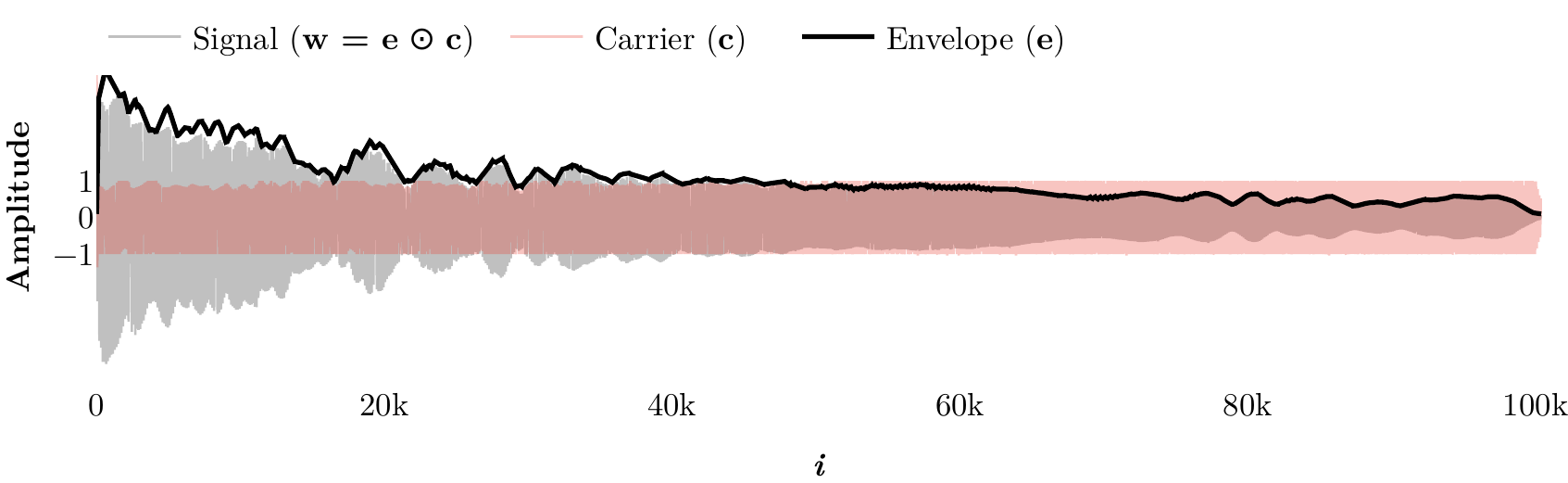}
  \caption{Signal, envelope and carrier for a record of a bend performed on an electric guitar.}
  \label{fig:Bend}
\end{figure}

In an effort to formalize the assessment of the quality of temporal envelopes, \textcite{1996Loughlinamplitude} proposed some conditions necessary, but not sufficient, to ensure the physical plausibility of an envelope. We comment briefly below that the algorithm here presented satisfies the four conditions presented.

The first one states that, if a signal is bounded in magnitude, then its envelope should be, as well. Our algorithm complies with this requisite, as the envelope is composed of selected samples from the original signal itself.

The second statement is that if the signal has a finite frequency range, that frequency range must not be exceeded by the envelope. This is only directly applicable in the continuous case, where this condition was proposed.

Nevertheless, if we picture both the wave and the envelope as samples from continuous functions, is easy to see that this is the case for our algorithm; since the envelope is composed by some samples of the digital wave, in the limit where all the samples of the wave are also part of the envelope, we have that the maximum frequency for both is the same. 

In all other cases, the maximum frequency of the envelope will be smaller than that of the original wave. The range of frequencies of the envelope, then, will always lie between zero and the maximum frequency of the original wave.

The third condition states what, for a periodic signal, the envelope should be a straight line with the intercept equal to the amplitude of this signal. Since all the local maxima have the same amplitude in the case of a periodic wave, our envelope will indeed be a straight line, with the same amplitude as the original wave. An example of such behaviour can be observed in figure \ref{fig:Comparison_sinusoid}.

The fourth and last condition states that if a wave is multiplied by a constant, the envelope should also be multiplied by the same constant. That is also the case, again because the points that form the envelope are points of the original wave itself.

\subsection{Comparison with traditional algorithms}

Direct comparison with many of the more recent algorithms is made difficult by the unavailability of digital implementations of such works, many designed to process analog signals \parencite[e.g.,][]{2018AssefModeling}.

Nevertheless, insight can be gained from a comparison of the results of the method here proposed with some of the most common envelope extraction algorithms. Specifically, we compare the present method with the following approaches:

\begin{enumerate}
\item Smoothing: This is a relatively simple procedure that, applied to the absolute values of a discrete signal, can provide an approximation of the variation of its amplitude in time. We use the Savitzky-Golay algorithm \parencite{1964SavitzkySmoothing} with a window width of 3001 samples and a cubic polynomial.
\item Filtering: The absolute value of the wave is filtered using a digital implementation of the Butterworth \parencite{butterworth1930theory} low pass filter. We use an order 2 filter with a cut off frequency of 10 Hz.
\item Hilbert Transform: We apply the Hilbert transform to the absolute values of the wave after filtering the original signal with a Butterworth filter with a cut off frequency of 100 Hz, using the absolute value of the analytic signal to obtain the envelope.
\end{enumerate}

\begin{figure}[H]
  \centering
    \includegraphics{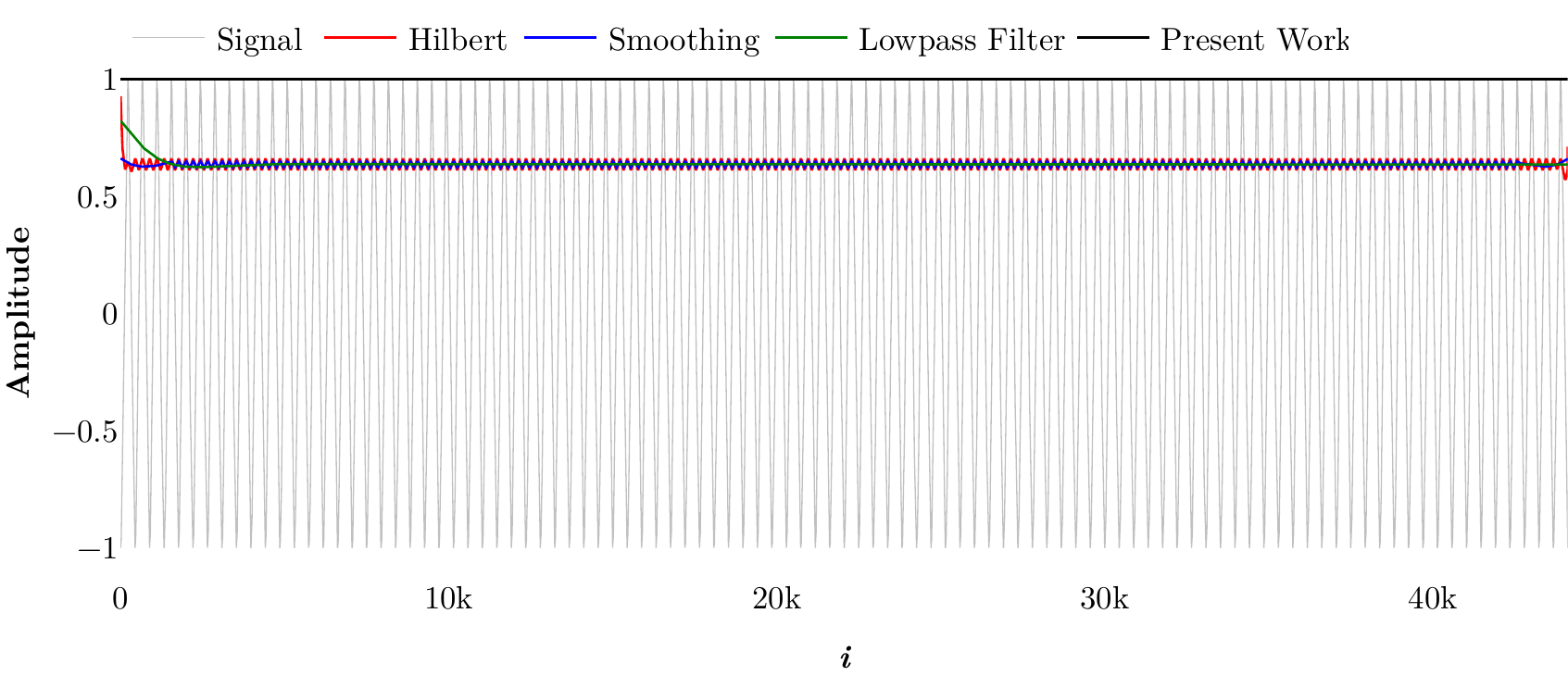}
  \caption{Comparison of envelope detection algorithms for a simple sinusoid}
  \label{fig:Comparison_sinusoid}
\end{figure}

In \ref{fig:Comparison_sinusoid} we illustrate the envelope extracted by the conventional algorithms, contrasting it with the result of the method presented in this work, for a simple sinusoid. Note that none of the conventional algorithms complies with the third condition proposed in \textcite{1996Loughlinamplitude} that the envelope of a periodic wave should be a straight line.

\begin{figure}[H]
  \centering
    \includegraphics{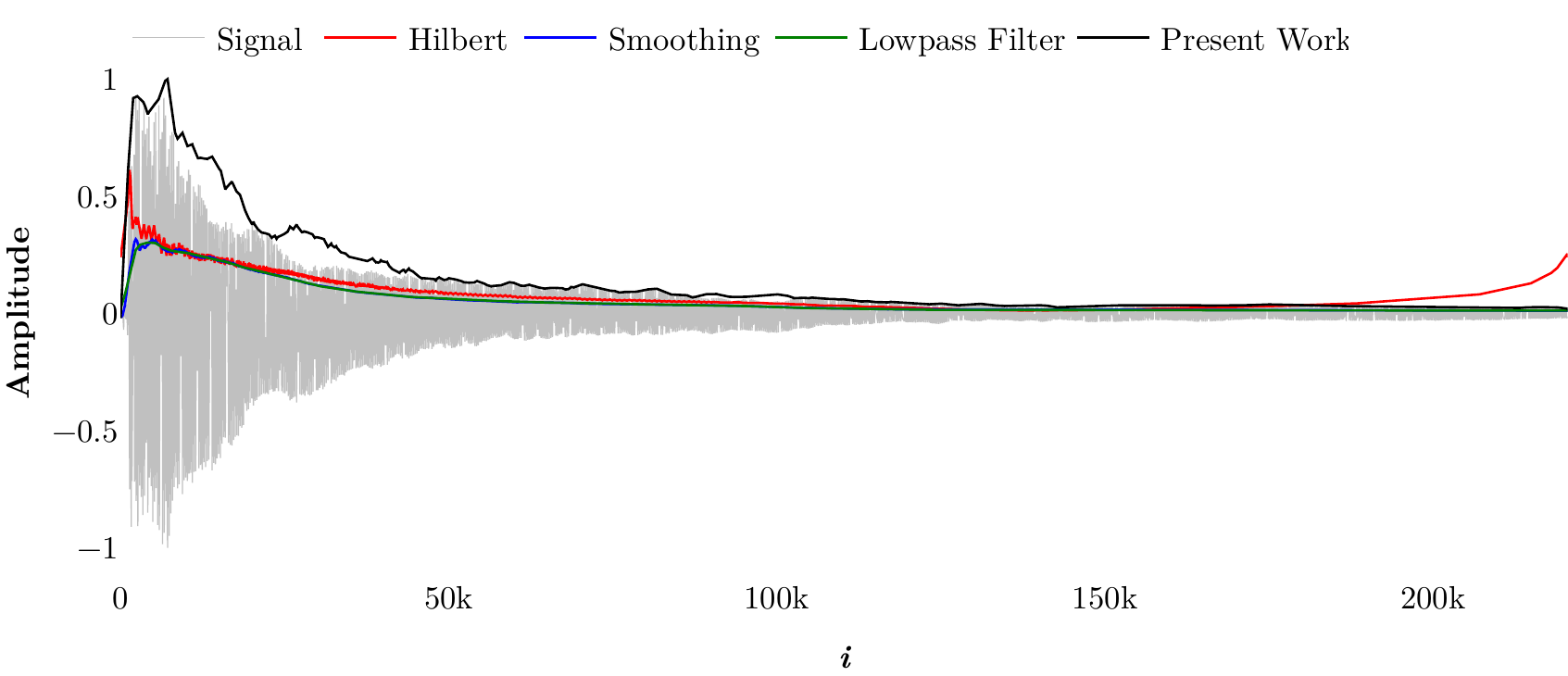}
  \caption{Comparison of envelope detection algorithms for the recording of the key 33 of a grand piano}
  \label{fig:Comparison_piano}
\end{figure}

For a recording of the key 33 of a grand piano, where parts of the sound are highly percussive and noisy, the Hilbert approach presents considerable oscillations, as can be seen in figure \ref{fig:Comparison_piano}.

\begin{figure}[H]
  \centering
    \includegraphics{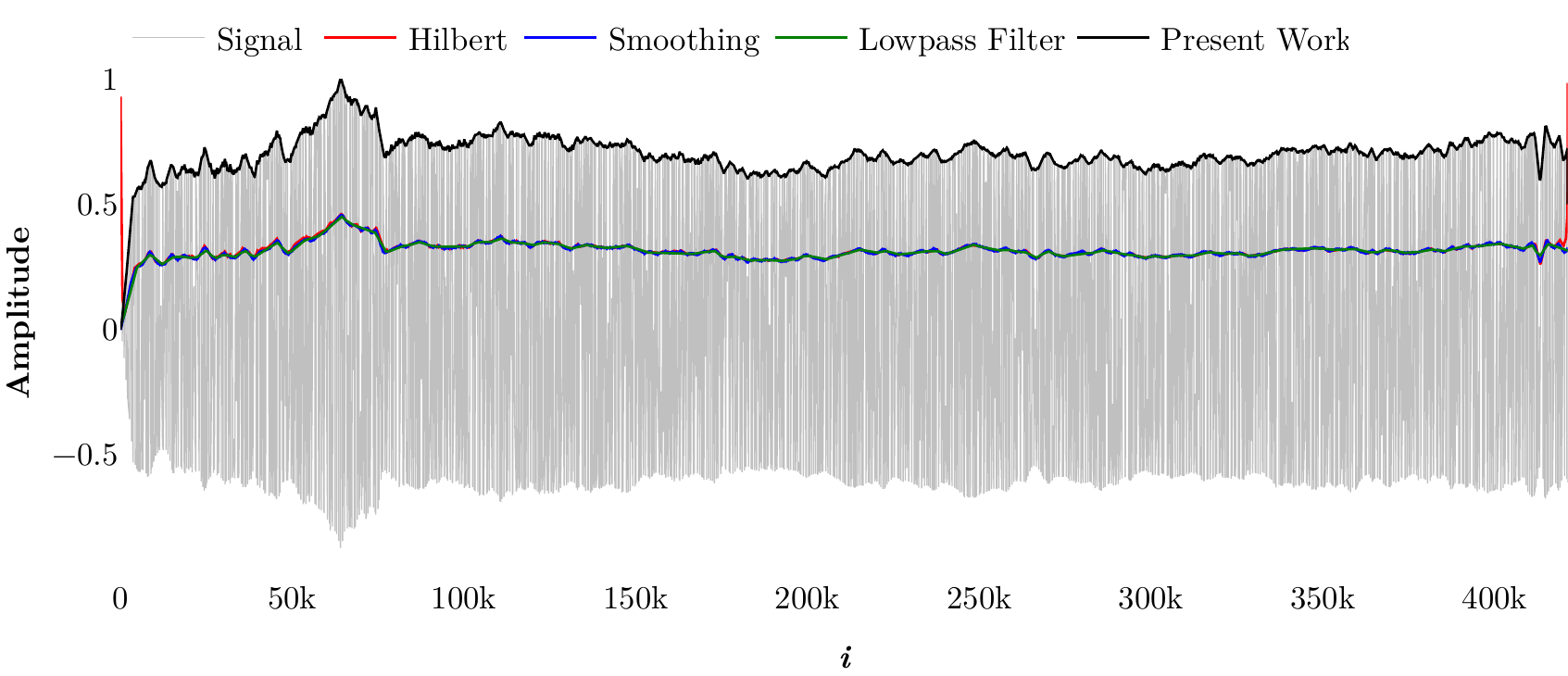}
  \caption{Comparison of envelope detection algorithms for the recording of the vocalization of a soprano singer}
  \label{fig:Comparison_soprano}
\end{figure}

In figure \ref{fig:Comparison_soprano} all envelopes exhibit a similar movement along the samples, but the ones generated by traditional algorithms are undershooting the original wave.

All the traditional methods demanded careful choice of parameters. In the case of the Hilbert transform, pre-processing in the form of filtering was applied. That is in contrast with our method, that organically tunes itself via the automatic choice of the radius of the circle representing the average discrete curvature of the signal.

Table \ref{table:Error_comparison} presents a comparison of the average errors of the techniques. The average error was calculated by the average of the squared distance of the points in the envelope divided by two and the absolute value of the wave. The rationale is that the values of the envelope divided by two should pass near the average of the absolute values of the wave. This metric is only valid for reasonably smooth envelopes, however.

We can see from the table \ref{table:Error_comparison} that the method presented here consistently outperforms the traditional approaches for all waves tested, yielding an error 33\% smaller, on average, than the other methods. It is also worth noting that the errors of the traditional methods are very similar between them.

\begin{table}[H]
\centering
\begin{tabular}{lcccc}
\hline
\multicolumn{5}{c}{\textbf{Average error per frame}} \\
              & Present Work & Smoothing & Low pass & Hilbert \\ 
\hline
alto          & 0.073        & 0.083     & 0.083   & 0.082   \\
bend          & 0.009        & 0.015     & 0.015   & 0.014   \\
piano         & 0.005        & 0.006     & 0.006   & 0.005   \\
brass         & 0.024        & 0.035     & 0.035   & 0.033   \\
soprano       & 0.042        & 0.065     & 0.065   & 0.065   \\
spoken voice  & 0.024        & 0.037     & 0.037   & 0.034   \\
tom           & 0.003        & 0.005     & 0.005   & 0.005   \\
sinusoid      & 0.113        & 0.193     & 0.195   & 0.191   \\ 
\hline
Average       & 0.037        & 0.055     & 0.055   & 0.054  
\end{tabular}
\caption{Comparison of the average errors of the algorithm presented in this work with the most common methods of digital envelope identification. The discrete waves used were normalized between -1 and 1. The average is between the absolute value of those waves and the envelope divided by two.}
\label{table:Error_comparison}
\end{table}

The times taken for the algorithms compared here to process each wave are shown in \ref{table:Time_comparison}. We used specialized implementations available in libraries of the Python programming language. The presented algorithm is in average the second-fastest.

\begin{table}[H]
\centering
\begin{tabular}{lcccc}
\hline
\multicolumn{5}{c}{\textbf{Time in seconds}} \\
              & Present Work & Smoothing & Low pass & Hilbert \\ 
\hline
alto          & 0.008        & 0.101     & 0.007   & 0.011   \\
bend          & 0.030        & 0.226     & 0.012   & 0.131   \\
piano         & 0.034        & 0.374     & 0.014   & 0.042   \\
brass         & 0.032        & 0.297     & 0.011   & 0.045   \\
soprano       & 0.177        & 0.735     & 0.027   & 0.566   \\
spoken voice  & 0.014        & 0.105     & 0.003   & 0.038   \\
tom           & 0.007        & 0.084     & 0.018   & 0.026   \\
sinusoid      & 0.007        & 0.091     & 0.005   & 0.011   \\ 
\hline
Average       & 0.039        & 0.252     & 0.012   & 0.109  
\end{tabular}
\caption{Comparison of the processing time of the algorithm presented in this work with the most common methods of digital envelope identification.}
\label{table:Time_comparison}
\end{table}

\subsection{Frontiers}

In practice, is not uncommon for a discrete wave, especially in the case of sound, to present somewhat different positive and negative contours; in those cases, the algorithm here presented can be used independently in the positive and negative pulses of the wave to define two “envelopes”, that we are going to call superior and inferior frontiers.

\begin{figure}[H]
  \centering
    \includegraphics{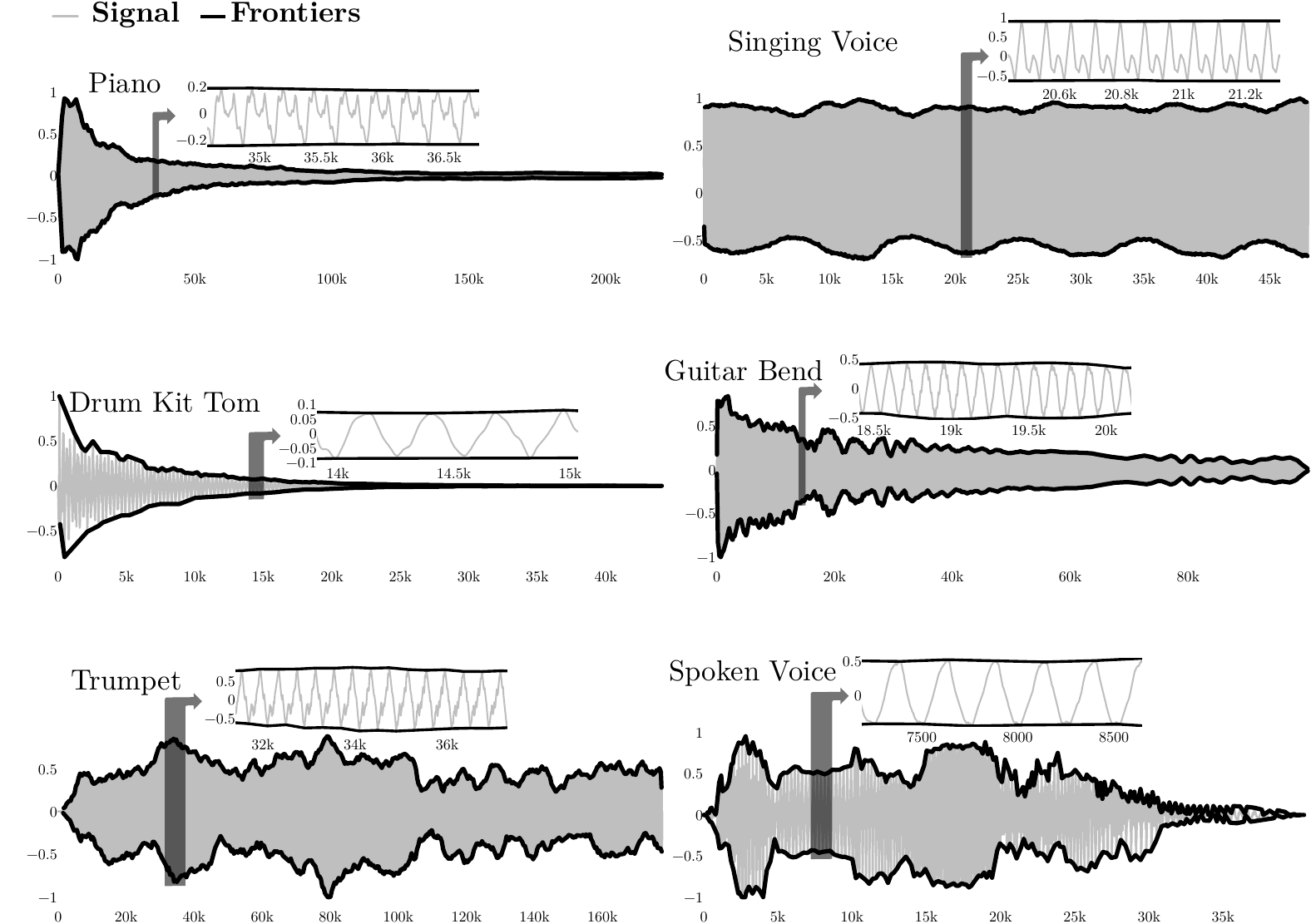}
  \caption{Positive and negative frontiers of six digital waves, as extracted by the algorithm here presented. For each wave, the region highlighted in black is shown in detail besides the whole wave. For each wave, the horizontal axis is the sample number $ i $, while the vertical axis is the normalized amplitude. All 6 waves were recorded at 44100 fps.}
  \label{fig:FullFrontiers}
\end{figure}

Figure \ref{fig:FullFrontiers} illustrates the frontiers of six diverse discrete sound waves, as well as an in-detail view of the highlighted segment for each signal. All waves are records of physical sounds, chosen to represent the applicability of the algorithm in real-world scenarios.

The frontiers are satisfactorily detected in the case of harmonic and inharmonic sounds, and are robust in relation to the number of samples and the frequencies of the waves. That is a peculiarity of our algorithm: the traditional algorithms are unable to generate two distinct envelopes for the same wave.

\subsection{Approximated location of pseudo-cycles}

\begin{figure}[H]
  \centering
    \includegraphics{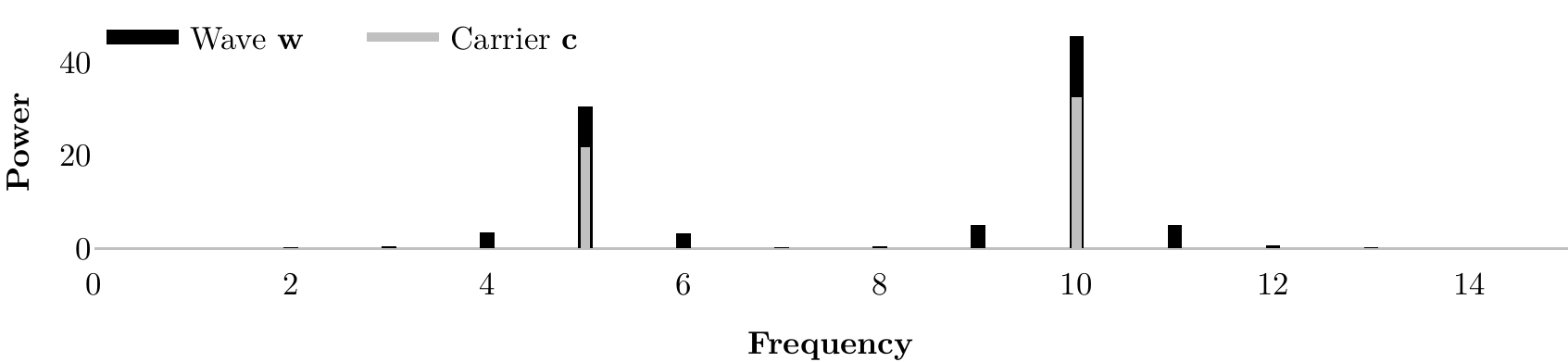}
  \caption{Fourier power spectrum for the wave and carrier shown in \ref{fig:Envelope}.}
  \label{fig:Fourier}
\end{figure}

The envelope is shown to add complexity to the spectral representation of a wave \parencite{2019TarjanoNeuro}, and an accurate description of the envelope, while describing the evolution of the instantaneous amplitude of a signal in time, would also simplify further spectral analysis. Figure \ref{fig:Fourier} shows the frequency-domain power spectrum for the wave and the carrier presented in \ref{fig:Envelope}. For the carrier, the power spectrum presents two nonzero values, while the power spectrum for the wave composed of the carrier modulated by the envelope is more complex.

Moreover, the algorithm developed in this work naturally divides a signal into its pseudo-cycles, pinpointing them in the time-domain, thus providing the building blocks for the reconstruction of the fine structure of the wave.

\begin{figure}[H]
  \centering
    \includegraphics{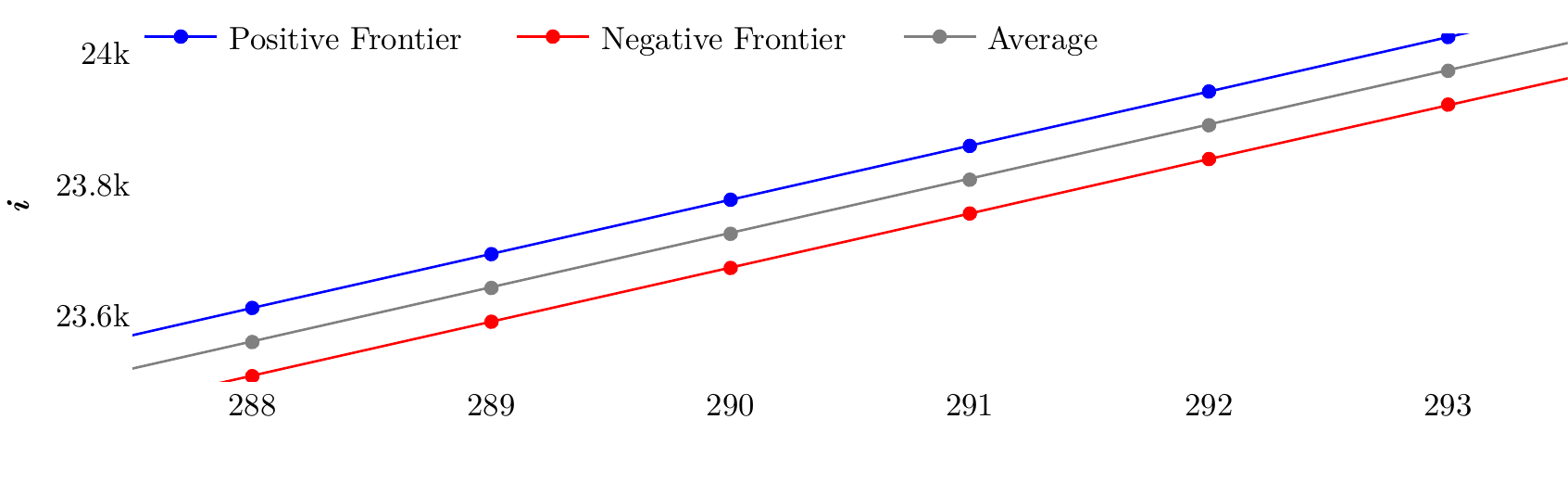}
  \caption{Section of the lines formed by the positions of the positive and negative frontiers of the digital wave illustrated in \ref{fig:singer_voice}, as extracted by the algorithm presented, and their average.}
  \label{fig:PCs}
\end{figure}

\begin{figure}[H]
  \centering
    \includegraphics{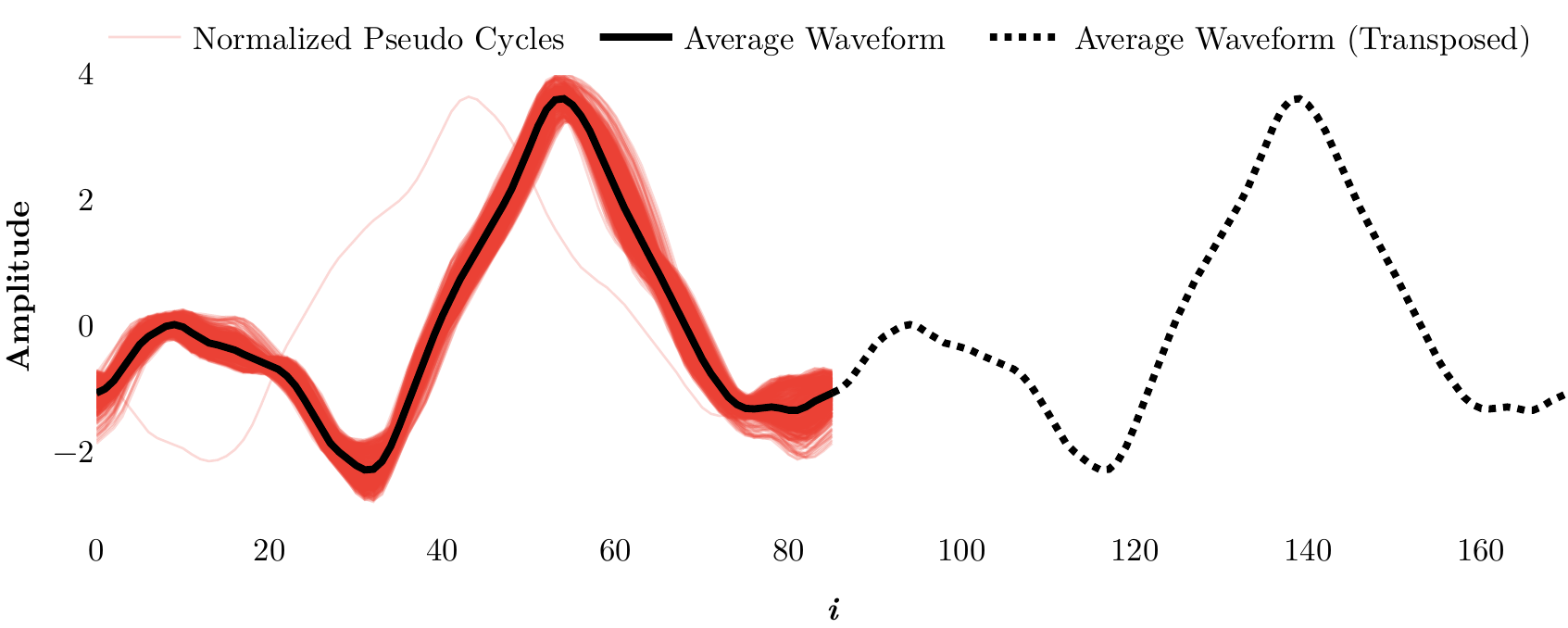}
  \caption{Pseudo-cycles of the wave illustrated in \ref{fig:singer_voice}, segmented and superposed, and their average. The average is also shown shifted by one average period, to illustrate the average continuity.}
  \label{fig:AvgPc}
\end{figure}

One can see in figure \ref{fig:PCs} a section of the plot of where the local maxima and minima occurs in the recording of an alto singer show in \ref{fig:singer_voice}. The average between the two is also shown. With this information, one can split a discrete wave into its pseudo-cycles, as is illustrated in Figure \ref{fig:AvgPc}: It shows the superposition of this extracted pseudo-cycles, after being normalized by length, and their average. This same average is shown shifted, to exemplify that approximate $C^0$ and $C^1$ continuity is achieved. 

\section{Conclusion}
This work fills a gap identified in the theory of digital signal processing, where the lack of general procedures for the accurate identification of the temporal envelope of arbitrary waves poses an obstacle to the complete description and eventual manipulation of signals.

By presenting a general approach for envelope detection, and eliminating the need for parameter tuning, we expect to facilitate further research at the many areas that rely on envelope detection techniques, and improve already existing algorithm that have envelope detection as an intermediary step.

Illustrating the efficiency of an algorithm inspired by geometric aspects of a discrete signal on a wide range of real-world signals, we also hope to encourage research in this direction.

The present work also highlights some gaps in our current understanding and definition of temporal envelope and, in so doing, endeavours to motivate further research in this topic, that can ultimately give rise to a more mathematically sound envelope theory, especially concerning broadband waves; suggestions in this direction are also presented.

While relevant in its own accord, the procedure here presented approximately isolates the individual pseudo-cycles of a wave, pinpointing them in the time-domain. 

Similarly, by decomposing a wave into its envelope and carrier, and simplifying the representation of such carrier in the frequency-domain, a rich set of advancements, especially in the areas of sound compression and synthesis in the time and frequency-domains, can be built upon this initial algorithm.

By illustrating how this information can be used to obtain the average waveform of a discrete wave, we hope to encourage research in this area, ultimately leading to alternative representations of discrete waves.

\section{References}

\printbibliography[heading=none]

\end{document}